\documentclass[12pt]{article}

\pdfoutput=1

\usepackage{graphicx}
\usepackage{epstopdf}
\usepackage[centertags]{amsmath}
\usepackage{amsthm}
\usepackage{amsfonts}
\usepackage{ccaption}
\usepackage[usenames]{color}
\usepackage{mathrsfs}

\usepackage[
      colorlinks=true,
      linkcolor=blue,
      urlcolor=blue,
      filecolor=blue,
      citecolor=red,
      pdfstartview=FitV,
      pdftitle={},
        pdfauthor={Veronika Hubeny, Don Marolf, Mukund Rangamani},
        pdfsubject={Black holes on the brane},
        pdfkeywords={black holes, Hawking radiation, deconfinement},
        pdfpagemode=None,
        bookmarksopen=true
      ]{hyperref}

\usepackage{rotating}
\makeatletter
\@addtoreset{equation}{section}

\makeatletter
\renewcommand\section{\@startsection {section}{1}{\z@}%
                                   {-3.5ex \@plus -1ex \@minus -.2ex}%nn
                                   {2.3ex \@plus.2ex}%
                                   {\normalfont\large\bfseries}}
\renewcommand\subsection{\@startsection{subsection}{2}{\z@}%
                                     {-3.25ex\@plus -1ex \@minus -.2ex}%
                                     {1.5ex \@plus .2ex}%
                                     {\normalfont\bfseries}}

  \captionnamefont{\bfseries}
  \captiontitlefont{\small\sffamily}
  \captiondelim{: }
  \hangcaption

\newenvironment{list2}{
  \begin{list}{$\star$}{%
      \setlength{\itemsep}{0.05in}
      \setlength{\parsep}{0in} \setlength{\parskip}{0.05in}
      \setlength{\topsep}{0in} \setlength{\partopsep}{0in}
      \setlength{\leftmargin}{0.4in}}}{\end{list}}

\newcommand{\be}{\begin{equation}}
\newcommand{\ee}{\end{equation}}
\newcommand{\beq}{\begin{eqnarray}}
\newcommand{\eeq}{\end{eqnarray}}

\def\sec#1{\S\ref{#1}}
\def\fig#1{Fig.\,\ref{#1}}
\def\req#1{(\ref{#1})}
\def\App#1{Appendix \ref{#1}}

\def\({\left(}
\def\){\right)}
\def\[{\left[}
\def\]{\right]}

\def\p{\partial}

\def\CB{{\cal B}}

\def\CE{{\cal E}}

\def\CM{{\cal M}}

\def\CR{{\cal R}}
\def\CS{{\cal S}}
\def\CT{{\cal T}}

\def\R{{\mathbb R}}

\def\A5S5{{\rm AdS}_5 \times \S^5}

\def\p{\partial}

\def\bit#1{ \noindent {\color{blue}{\it #1}}}

\def\H{{\bf H}}

\def\AdS#1{AdS$_{#1}$}
\def\SAdS#1{Schwarschild-AdS$_{#1}$}

\title{{\bf \Large Black funnels and droplets from the AdS C-metrics}}

\author{\normalsize
Veronika E. Hubeny$^{a,b}$\footnote{veronika.hubeny@durham.ac.uk},\ \ Donald Marolf$^{\,c\,}$\footnote{marolf@physics.ucsb.edu}, and
Mukund Rangamani$^{a,b}$\footnote{mukund.rangamani@durham.ac.uk} \\ \\
$^a$\small \sl  Centre for Particle Theory \& Department of
Mathematical Sciences,\\[-1.5mm]
\small \sl Science Laboratories, South Road, Durham DH1 3LE, United Kingdom. \\[1mm]
$^b$\small\sl Kavli Institute for Theoretical Physics, UCSB, Santa Barbara, CA 93015, USA.\\[1mm]
$^c$ \small \sl Physics Department, UCSB, Santa Barbara, CA 93106, USA.
}
\begin{document}

\setlength{\baselineskip}{16pt}
\begin{titlepage}
\maketitle
\begin{picture}(0,0)(0,0)
\put(350, 320){DCPT-09/59}
\put(350,305){NSF-KITP-09-165}
\end{picture}
\vspace{-36pt}

%Abstract
\begin{abstract}
We recently argued that the dynamics of strongly coupled field theories in black hole backgrounds is related via the AdS/CFT correspondence to two new classes of AdS black hole solutions: black funnels, and black droplets suspended above a second disconnected horizon.   The funnel solutions are dual to black holes coupling strongly to a field theory plasma.  In contrast, the droplet solutions describe black holes coupling only weakly.  We continue our investigation of these solutions and construct a wide variety of examples from the AdS C-metric in four bulk spacetime dimensions. The solutions we find are dual to field theories on spatially compact universes with Killing horizons.
 \end{abstract}
\thispagestyle{empty}
\setcounter{page}{0}
\end{titlepage}

\renewcommand{\thefootnote}{\arabic{footnote}}
%______________________________________

%%%%%%%%%%%%%%%%%%%%%%%%%%%%%%%%%%%%%%%%%%%%

\tableofcontents

%~~~~~~~~~~~~~~~~~~~~~~~~~~~~~~~~~~~~~~~~~~~~~~~
\section{Introduction}
\label{s:intro}
%~~~~~~~~~~~~~~~~~~~~~~~~~~~~~~~~~~~~~~~~~~~~~~

The AdS/CFT correspondence \cite{Maldacena:1997re} provides a unique window into the dynamics of a class of strongly coupled gauge field theories. For large $N$ gauge theories, the dynamics in the planar limit is  expected to be effectively classical, with $1/N$ controlling the quantum corrections. For a class of superconformal field theories arising as world-volume theories on D-branes or M-branes, the AdS/CFT correspondence identifies this classical dynamics of the single trace sector with that of classical string theory in a higher dimensional spacetime.  Furthermore, if the field theory is strongly coupled then one can truncate to the zero mode sector of the string theory, viz., classical gravity in this higher dimensional spacetime. The correspondence therefore provides an avenue to explore the strong coupling dynamics of field theories by reformulating the physics in terms of an effective classical  gravity theory.

In this paper we continue our investigation of strongly coupled field theories on black hole backgrounds using the AdS/CFT correspondence, generalizing the results of \cite{Hubeny:2009hr}. Field theories in curved spacetime are known to exhibit a rich array of physical phenomena ranging from vacuum polarization and particle production to Hawking radiation and its associated puzzles with information loss. However, much of the investigation in the past has focussed on perturbative field theory due to the lack of access to the full non-perturbative quantum state, even in the context where gravity is non-dynamical. Our current interest lies in understanding relevant quantum states beyond perturbation theory; this is where AdS/CFT comes into play.

Consider a field theory on a non-dynamical curved spacetime (which we denote as $\CB_d$) with metric $\gamma_{\mu\nu}$. We would like to know the behavior of interesting quantum states and in particular the expectation values of gauge invariant local operators at the non-perturbative level.  If we restrict attention to strongly coupled conformal fields which arise in low-energy limits of D-brane world-volume theories, then we can exploit the AdS/CFT correspondence to answer these questions. This is achieved by identifying higher-dimensional asymptotically AdS gravitational solutions dual to the desired field theory states on $\CB_d$; we will refer to these gravitational saddle points as $\CM_{d+1}$. Such bulk spacetimes are found by solving the gravitational equations of motion subject to the boundary condition that  $\CM_{d+1}$ has as its  timelike boundary $\CB_d$.\footnote{Generically the correspondence only requires that the boundary $\p \CM_{d}$ of the bulk spacetime $\CM_{d+1}$ be in the same conformal class as $\CB_d$. We will however demand that $\p \CM_{d}$ in fact is isometric to $\CB_d$.}  In particular, smooth static spacetimes are candidate duals for the field theory Hartle-Hawking states.    Of course,  it might turn out that the field theory in question has a non-trivial phase structure, which implies that there are multiple such static saddle points for the bulk gravity description \cite{Witten:1998zw}.

\begin{figure}[h]
\begin{center}
\includegraphics[scale=1.1]{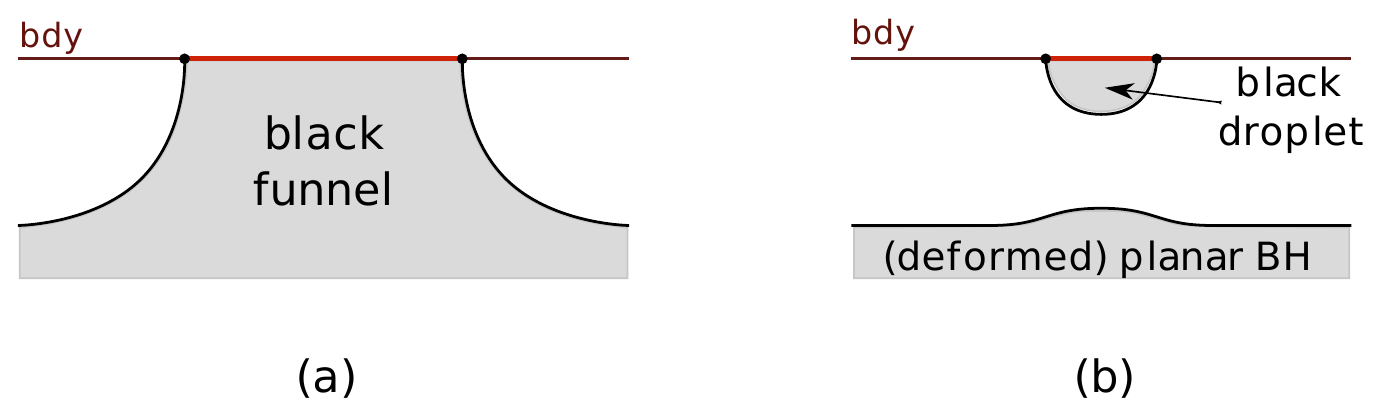}
 \caption{A sketch of our two novel classes of solutions: {\bf (a):} black funnel and  {\bf (b):} black droplet above a deformed planar black hole. }		
\label{f:fundrop}
\end{center}
\end{figure}

In \cite{Hubeny:2009hr} we examined in some detail the holographic duals of field theories on black hole backgrounds $\CB_d$. In particular, we have argued that there are new classes of black hole geometries in asymptotically AdS spacetimes: i) single connected horizon solutions which we called black funnels and ii) solutions with two disconnected horizons; see \fig{f:fundrop}. In the latter case, we called the component connected to the boundary horizon a black droplet.\footnote{In principle, at least for non-conformal theories, droplets can also exist without a second horizon being present.  In this case, a droplet is distinguished from a funnel by its behavior far from the boundary black hole.  In particular, droplet horizons must be compact with respect to the conformally rescaled metric which asymptotes to $\gamma_{\mu\nu}$ on the boundary (we will refer to this as ``$\gamma$-compact").  In contrast, when the boundary spacetime has a good asymptotic region describing physics far from the boundary black hole, a black funnel should asymptote to the bulk solution describing a deconfined plasma in this distant region of spacetime. For example, in the case of asymptotically flat boundaries, it should asymptote to the planar AdS black hole.  As we discuss below, the distinction between droplets and funnels is more subtle for spatially compact boundary metrics.} The funnel solutions are dual to black holes coupling strongly to the field theory plasma. In contrast, the droplet solutions describe black holes coupling only weakly.  In particular, the second outer horizon present in these solutions is interpreted as the field theory plasma, while the droplet itself describes field theory vacuum polarization near the horizon.  The lack of connection between these two AdS horizons is evidence of the weak coupling.   Note that while the gravitational dynamics may allow solutions where the two horizons have different temperatures, only the equal temperature solutions can be dual to field theory Hartle-Hawking states.

As evidence for this picture, ref.\ \cite{Hubeny:2009hr} exhibited funnel and droplet solutions dual to $1+1$ and $2+1$ dimensional conformal fields living on black hole backgrounds.   In the $1+1$ case, we constructed  black funnel solutions in \AdS{3} whose boundary is the two dimensional black hole \cite{Witten:1991yr}.  While black droplets do not arise for $1+1$ boundary black holes,
by exploiting the known AdS C-metric solutions we were also able to construct both black funnels and black droplets in four bulk spacetime dimensions.  However, the droplet solutions found in this way do not generically describe Hartle-Hawking states as they were accompanied by a second horizon of a different temperature, so that the solutions did not describe equilibria.  Due to the properties of the C-metric, the relevant 2+1 boundary metrics described black holes which asymptote to $\R \times \H^2$.

A natural question to ask is whether further interesting solutions are hidden among the AdS C-metrics. In the present work, we analyze this issue within the family of uncharged, non-rotating AdS C-metric solutions.  Recall that the AdS C-metric solutions have been useful in the past to
construct localized black holes on a UV brane in the brane-world context \cite{Emparan:1999wa, Emparan:1999fd} and also more recently to construct plasma ball solutions on an IR brane \cite{Emparan:2009dj} (see also \cite{Anber:2008qu,Anber:2008zz}) Our interest is to remove the UV brane and work with some prescribed boundary metric. Since we are not a-priori fussed about what metrics we have on the boundary (apart from the fact that they be black hole like), it seems plausible that new interesting solutions will emerge.  As we shall see in the following, there is indeed a rich class of boundary black holes contained within the C-metric family.

With this motivation,  we undertake an exhaustive search of the AdS C-metric family of solutions and find an interesting class of black funnel and black droplet solutions.  In all cases, by using standard holographic methods we are able to compute the boundary stress tensor which includes the contribution from the quantum dynamics of the field theory in curved spacetime. We find that the stress tensor does indeed capture the thermal aspect of Hawking radiation and is furthermore regular on the black hole horizon in the boundary.

On the boundary we generally find black holes living in spatially compact universes; i.e., there are no spatial asymptopia.   This class of solutions is therefore somewhat different from those which arise for the special choice of parameters examined in \cite{Hubeny:2009hr}, where the boundary metric was asymptotically a hyperbolic cylinder $\R \times \H^2$.  Due to the absence of an asymptotic region, the precise definitions of funnel and droplet given in   \cite{Hubeny:2009hr} do not apply.  We therefore extend these definitions in  \sec{s:coordcmet} below.  Our new definitions are sufficient for spacetimes such as the C-metric which have an appropriate rotational Killing field, even if they lack a useful asymptotic region.

The organization of this paper is as follows: we begin in \sec{s:adscmet} with a brief overview of the AdS C-metric solutions. While these geometries have been studied in the literature before, we find it useful to review and generalize some of the results, especially those pertaining to the precise coordinate domains.  We then analyze the C-metric family in detail in section \sec{s:coordcmet}, where we show that, apart from trivial cases that are exactly AdS or a quotient, any uncharged, non-rotating asymptotically AdS C-metric with vanishing NUT charge can be interpreted in terms of funnels, droplets, and planar black holes.
The discussion can clearly be generalized to include additional charges, but we refrain from doing so here in order to keep the discussion simple.  We extract the boundary stress tensor for these solutions in \sec{s:cbdystress} which allows us to see the advertised thermal behavior of the field theories in black hole backgrounds.  We end with a discussion in \sec{s:discuss} and describe some subtle limits in Appendix \ref{s:lambn1}.

%~~~~~~~~~~~~~~~~~~~~~~~~~~~~~~~~~~~~~~~~~~~~~~~
\section{The AdS C-metric}
\label{s:adscmet}
%~~~~~~~~~~~~~~~~~~~~~~~~~~~~~~~~~~~~~~~~~~~~~~

The C-metric solution in four dimensions corresponds physically to a pair of black holes being uniformly accelerated by a cosmic string. The most general solution was found in \cite{Plebanski:1976gy} in the context of Einstein-Maxwell theory with a cosmological constant.  The general solution is specified by seven parameters, corresponding to the mass, angular momentum, an acceleration parameter, electric and magnetic charges, cosmological constant and a NUT parameter. We will be interested in a sub-class of these solutions which we will exploit in the context of the AdS/CFT correspondence to study four dimensional bulk spacetime duals of $2+1$ dimensional field theories living on a black hole background.

Consider the sub-class of AdS C-metrics \cite{Plebanski:1976gy}  whose line element is given in conventional C-metric coordinates as:\footnote{We have rescaled the timelike Killing field by a constant factor relative to the form of the C-metric used in \cite{Emparan:1999fd}.}
\begin{equation}
ds^2 = \frac{\ell^2}{(x-y)^2} \, \left( -\frac{F(y)}{1+\lambda}\, dt^2 + \frac{dy^2}{F(y)} + \frac{dx^2}{G(x) } + G(x)\, d\phi^2\right) \ ,
\label{adsc}
\end{equation}	
where  the functions $F$ and $G$ in \req{adsc} are of the form:
\begin{equation}
F(\xi) = \lambda +\kappa\, \xi^2 +  2\,\mu\, \xi^3  \ , \qquad G(\xi)  = \lambda + 1 - F(\xi) = 1 -\kappa\, \xi^2 - 2 \,\mu \, \xi^3 \ .
\label{FGdefs}
\end{equation}	
These metrics describe uncharged, non-rotating solutions with vanishing NUT charge which are negatively curved, i.e., they solve Einstein's equations with negative cosmological constant
\begin{equation}
\CE_{\mu\nu} = R_{\mu \nu} + \frac{3}{\ell_{4}^2} \, g_{\mu \nu} = 0 \ .
\label{eeddim}
\end{equation}	
%
%------
The AdS C-metric describes the geometry of accelerating black holes in an asymptotically AdS spacetime. In the metric \req{adsc}, $\ell$ captures the (inverse) acceleration, while $\lambda$ is related to the cosmological constant and $\mu$ is the mass parameter of the black hole(s).  Since cases with $\mu =0$ are locally isometric to either flat space or AdS${}_4$, we use the symmetry $x \to -x, y \to -y, \mu \to -\mu$ to take $\mu > 0$.
 The bulk AdS scale is
\begin{equation}
\ell_4 = \frac{\ell}{\sqrt{\lambda +1}},
\label{ads4cscale}
\end{equation}	
so that $\lambda \to -1$ at fixed $\ell$ is the flat space limit (and $\lambda <-1$ would give deSitter-C metrics).  We therefore take $\lambda > -1$.  Finally, $\kappa$ is a discrete variable taking values $\pm1, 0$ and corresponds to different allowed topologies for the black holes; $\kappa =1$ corresponds to topologically spherical horizons while $\kappa = -1, 0$ corresponds to non-compact horizons with $\R^2$ topology.

A detailed discussion of the AdS-C metric properties can be found in \cite{Emparan:1999wa,Dias:2002mi} for $\kappa =1$ and in \cite{Emparan:1999fd} for other values of $\kappa$.  Here we note only that by taking a suitable $\ell \to \infty$ limit one can recover the standard \SAdS{4} black hole for $\kappa =1$.  For $\kappa =-1$ one obtains the topological black hole of \cite{Emparan:1999gf} and one can get the planar AdS black hole in the case when
$\kappa=0$.

%~~~~~~~~~~~~~~~~~~~~~~~~~~~~~~~~~~~~~~~~~~~~~~~
\subsection{The geometry of the AdS C-metric}
\label{s:geocmet}
%~~~~~~~~~~~~~~~~~~~~~~~~~~~~~~~~~~~~~~~~~~~~~~

\begin{figure}[tp]
\begin{center}
\includegraphics[scale=0.9]{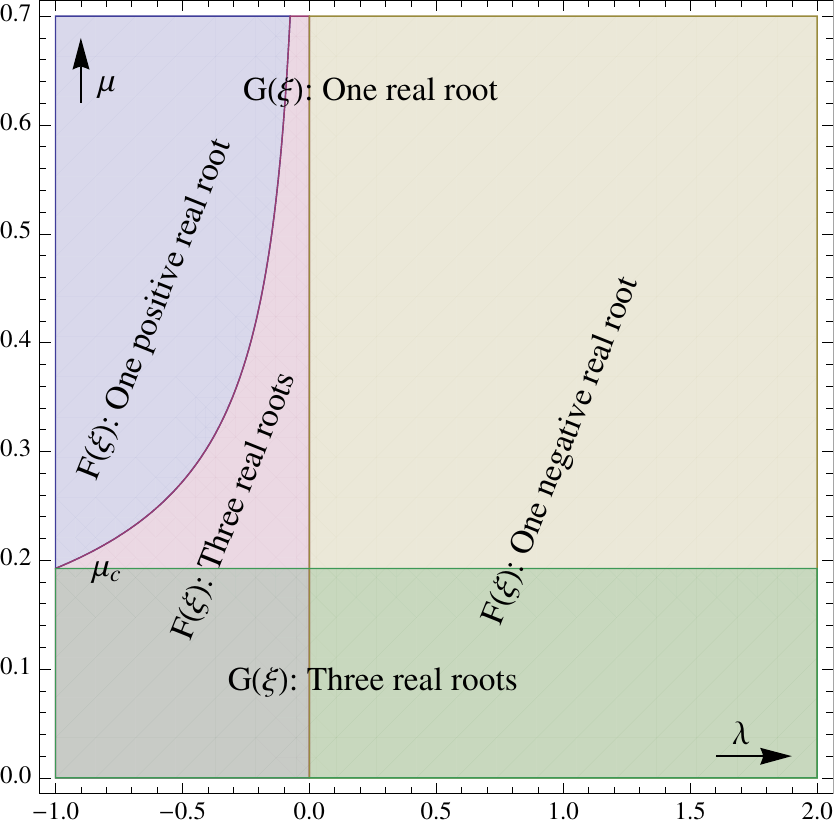}
  \caption{A plot of the domains in the $\{\lambda, \mu\}$ plane  which characterize the distinct possibilities for the root structure of $F(\xi)$ and $G(\xi)$ for $\kappa =1$. The behavior of $G(\xi)$ is simply controlled by the parameter $\mu$, while $F(\xi)$ has non-trivial behavior across the various domains as indicated.  See main text for a detailed explanation.}	
\label{f:kappa1dom}
\end{center}
\end{figure}

Let us now examine the geometry of the AdS C-metric \req{adsc}, \req{FGdefs} more carefully, to classify all the distinct possibilities as we vary the parameters $\lambda$, $\mu$, and $\kappa$ (we can ignore the parameter $\ell$ as it merely provides an overall scale).  The key features (boundaries, horizons, singularites) are determined by the coordinate ranges and the roots of the functions $F$ and $G$ in \req{FGdefs}.  Below, we first discuss the coordinate ranges and motivate the physical properties of the solutions, and in \sec{s:coordcmet} we study the root structure in more detail (summarized in \fig{f:kappa1dom} for the case of $\kappa=1$ and \fig{f:kappam1dom} for $\kappa=-1$).

Due to the conformal factor $(x-y)^{-2}$ in the metric, it is clear that the boundary of the spacetime is at $x=y$. Furthermore, the spacetime has singularities at $ y = \pm \infty$ and at $x = \pm\infty$ which are genuine curvature singularities; the Kretschmann scalar $R_{\mu\nu\rho \sigma}\, R^{\mu\nu\rho\sigma}$ diverges as $(x-y)^{6}$ \cite{Dias:2002mi}. Typically this has led previous analysis of the AdS C-metric to restrict attention to the region $-\infty < y \le x$. However, we will see that it is also sensible to consider the region $x \le y < \infty$, at least for certain choices of parameters. Of course, in making these choices we have to ensure that the spacetime in question doesn't have any naked singularities. We will return to this issue after a short examination of the coordinate ranges.

\begin{figure}[t!]
\begin{center}
\includegraphics[scale=0.82]{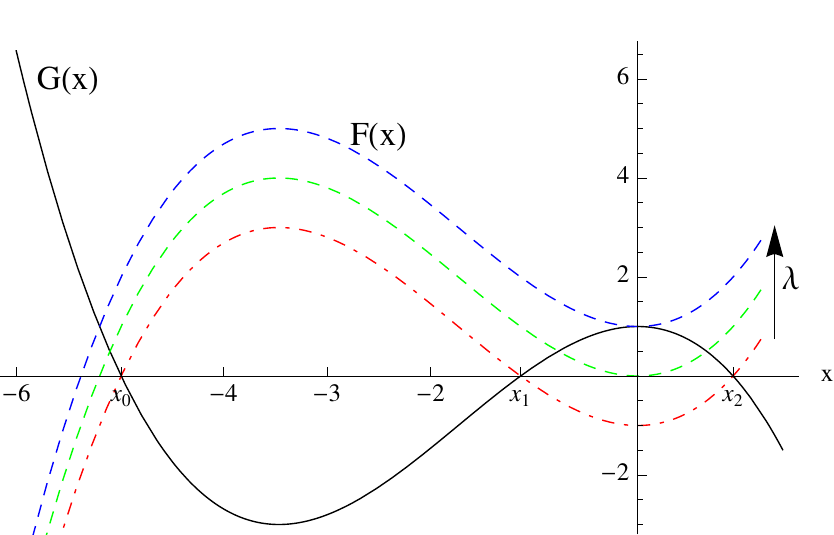}  \hspace{1cm}
\includegraphics[scale=0.82]{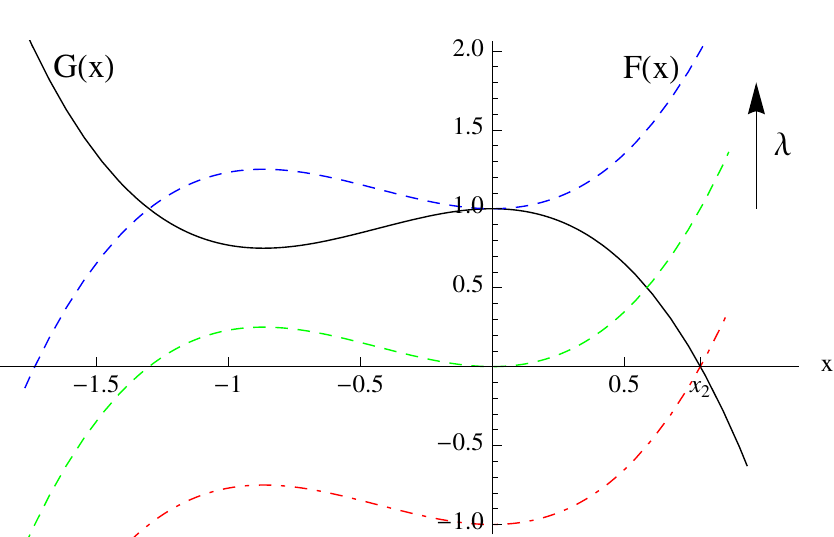} \\
\vspace{1.4cm}
\includegraphics[scale=0.82]{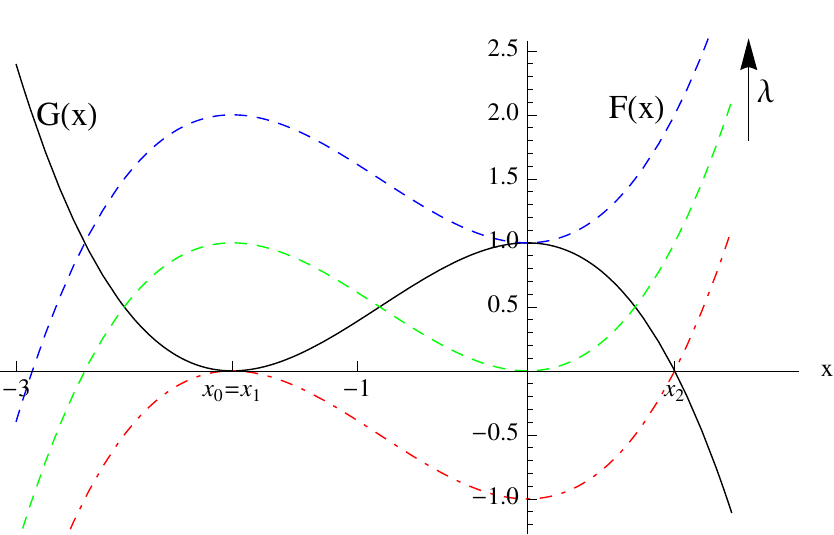} \\
\vspace{1cm}
\includegraphics[scale=0.82]{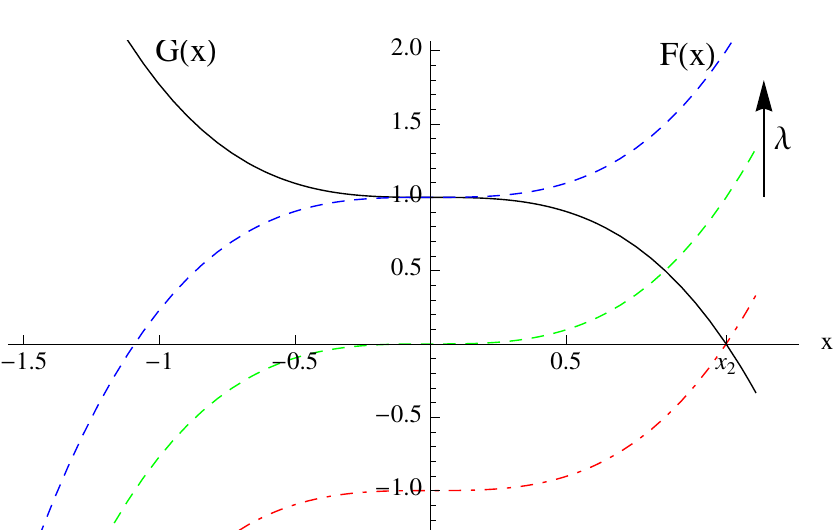}\hspace{1cm}
\includegraphics[scale=0.82]{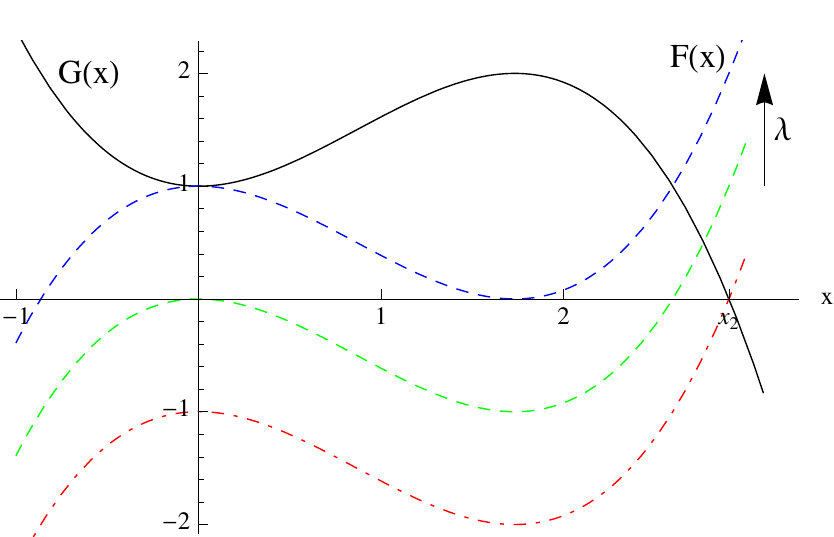}
\begin{picture}(0,0)
\setlength{\unitlength}{1cm}
\put(-13,10.7){$\kappa =1,  \; \mu = \frac{1}{2} \, \mu_c < \mu_c$}
\put(-5.5,10.7){$\kappa =1,  \; \mu = 2\mu_c > \mu_c$}
\put(-8.,5){$\kappa =1, \; \mu = \mu_c$}
\put(-12.5,-0.5){$\kappa =0$}
\put(-5,-0.5){$\kappa = -1$}
\end{picture}
\end{center}
  \caption{We plot the functions $F(x)$ and $G(x)$ for various values of $\lambda$ with the different panels corresponding to varying $\mu$ and $\kappa$ as indicated under the respective plots. The solid curve is the function $G(x)$ while the dashed curves correspond to $F(x)$ which are plotted for various values of $\lambda$ (which corresponds to the value of $F(0)$). The lowest (dot-dashed) curve has the limiting value $\lambda = -1$.   }	
\label{f:fgplots}
\end{figure}
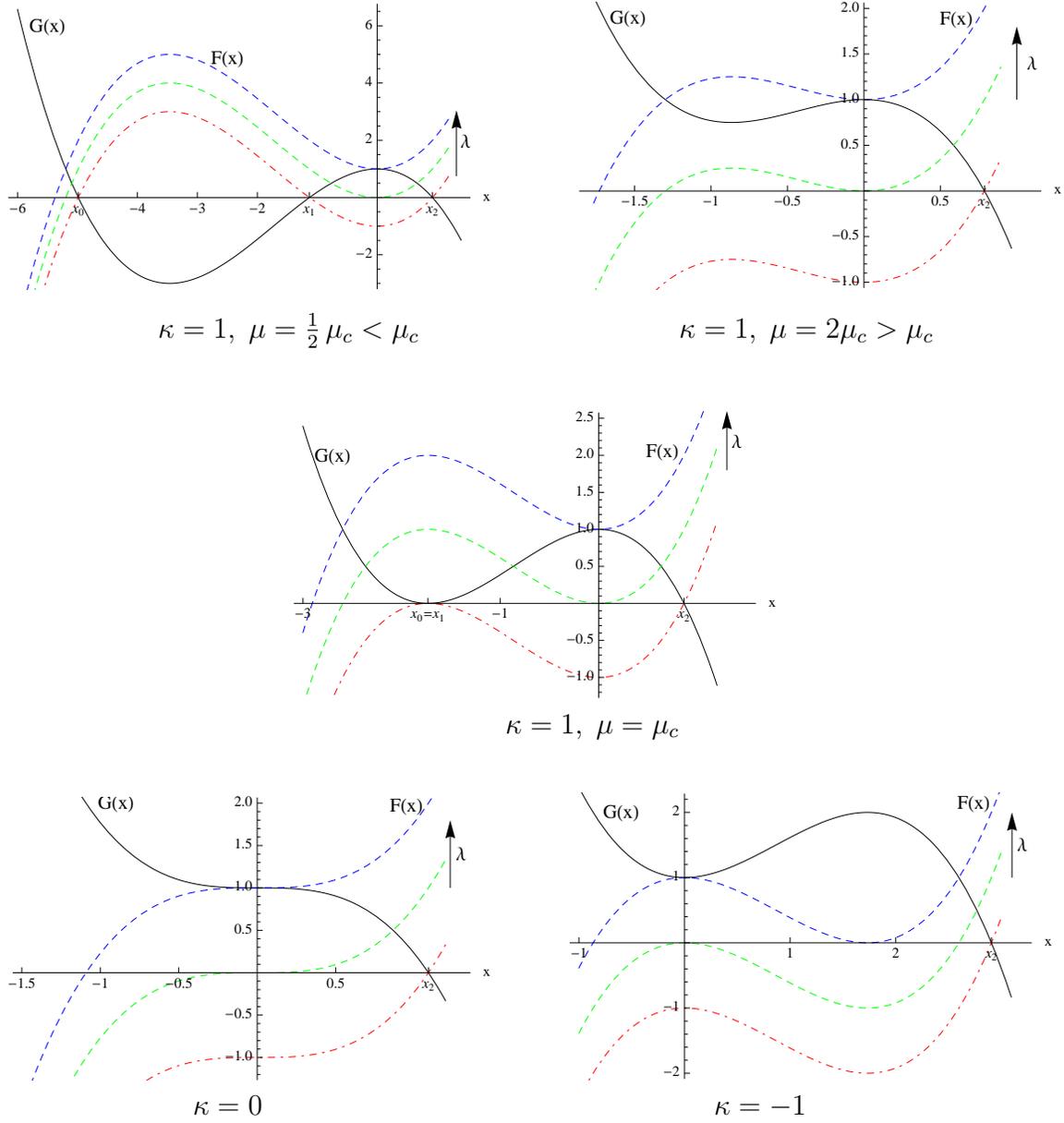

To determine the range of $x$ we need to examine the function $G(x)$. Being a cubic, $G(x)$ generically has three roots and as a result we can have two cases: (i) either the roots are all real, (ii) or only one root is real.
When $\kappa =1$ the distinction between the two situations is controlled by the mass parameter $\mu$, whereas for $\kappa=0,-1$ only the case (ii) occurs (see \fig{f:fgplots}).
We will order the roots as $x_0$, $x_1$ and $x_2$, with $x_0$ taken to be the smallest root of $G(x)$ in case (i) and we take $x_2$ to be the solitary root in case (ii).

Horizons arise at the values of $y$ for which $F(y)=0$.  We will denote the roots again by  $y_0$, $y_1$ and $y_2$ with $y_0 < y_1 < y_2$. In the situations where $F(y)$ has a single root, we will for simplicity denote it by $y_0$
(even when it is continuously connected to the $y_2$ root).
For instance,  in the simple case $\lambda = 0$ with $\kappa =1$ we have
\begin{eqnarray}
y_0 &=&  -\frac{1}{2\mu}  \ , \qquad \qquad \;\;\text{black hole horizon} \nonumber \\
y_1 &=& y_2 = 0 \ , \qquad \qquad \text{Poincar\'e horizon}
\label{}
\end{eqnarray}	

In general the formula for the roots $y_i$  as a  function of $\mu$ and $\lambda$ is messy and we will not write it down.  However, it is useful to note that from $\mu > 0$, the relation $F(\xi) + G(\xi) = 1+ \lambda > 0$, and the fact that $\xi=0$ is an extremum of both functions, one can deduce the ordering
\begin{equation}
\label{order}
y_0 < x_0 \le x_1 < y_1 \le 0 \le  y_2 < x_2
\end{equation}
when all roots are real.  See \fig{f:fgplots} for plots of the functions $F$ and $G$ for various values of parameters.   When some roots are complex, the remaining real roots generally still satisfy (\ref{order}) with the missing roots removed from the list.  A rather trivial exception occurs for $\kappa =1, \lambda = - \frac{1}{27 \mu^2}$ in which case we have $y_0 =  y_1 \le 0 \le  y_2 < x_2$, replacing two of the inequalities in (\ref{order}) by equalities.
The only other exception occurs in certain cases where $F$ and $G$ both have only one root for which $0 < y_0 < x_2$; i.e., only the relative order of $y_0$ and $0$ differs from (\ref{order}).   In the limit $\lambda \to -1$, the roots of $F$ and $G$ coincide and satisfy $y_0 = x_0 \le x_1 = y_1  < 0 < y_2 = x_2$
for $\kappa=1$ or $0<y_0<x_2$ for $\kappa=0,-1$.

% Figure

The temperature of any horizon is easily computed by noting that the Euclidean metric is regular when we identify the thermal circle  with period given by the inverse temperature $T_i^{-1}$ for
\begin{equation}
T_i = \frac{|F'(y_i)|}{4\pi\, \sqrt{1+\lambda}}  \ , \qquad {\rm for} \; i = 0,1,2\, .
\label{tphys}
\end{equation}	
Below, we will allow $x$ to range over intervals of the form $[x_{min},x_{max}]$ where $x_{max}$ is a root of $G$.  As a result, $||\partial_t || = -1$ at $y = x = x_{max}$ and our temperature corresponds to a unit normalized Killing field at $x_{max}$.

Before proceeding further let us also record the boundary metric given by setting $x=y$ and stripping off two powers of a conformal factor, which we choose to be $(x-y)/\ell$.  The result is
\begin{equation}
ds^2_{\text{bdy}} = -\frac{F(x)}{1+\lambda}\, dt^2 + \frac{(1+\lambda)\, dx^2}{F(x)\, G(x)} + G(x)\, d\phi^2.
\label{indbdymet}
\end{equation}	
This is the metric of a black hole with Killing horizons at $x = y_i$. Curiously, as we move away from the horizon in the range $y_k < x < x_i$, where $x_i$ is the root of $G(x)$ immediately to the right of $y_k$, we find that the size of the Euclidean time circle grows while that of the spatial $\phi$ circle shrinks. In particular, $G(y_i) = 1+\lambda$, $\forall \,i$,  indicating that the $\phi$ circle always has finite size at the horizon.

The AdS C-metric solution  can be  physically visualized in terms of a black hole pulled  by a cosmic string  which accelerates it. Once one identifies the location of the cosmic string one can ask whether it hits the boundary. The place where it hits the boundary is  a conical defect on the boundary and ideally one would like to hide this behind the black hole horizon or to choose the period of $\phi$ so that the defect disappears.  One also wishes to ensure that the bulk curvature singularities at $x,y = \pm \infty$ are likewise hidden behind horizons.

To probe this issue we need to consider all possible coordinate regions. We will do so in \sec{s:coordcmet} below, but let us first record two important facts which will play a crucial role in our analysis. By examining the behavior of the spatial part of the metric near a simple root of $G(x)$, say $x_i$, we learn that the spacetime will be regular provided we identify the coordinate $\phi$ with period
\begin{equation}
\Delta \phi = \frac{4\pi}{|G'(x_i)|}.
\label{phiperiod}
\end{equation}	
Secondly, depending on the range of the coordinates we choose, we will be interested in the proper distance between interesting points such as the event horizon and the location of the cosmic string on the boundary. This is easy to compute using the induced boundary metric
\req{indbdymet} and it is straightforward to see that for $x \in \left[x_{\text{min}} , x_{\text{max}}\right]$ one has
\begin{equation}
x_{\text{proper}} =  \sqrt{1+\lambda} \, \int_{x_{\text{min}}}^{x_{\text{max}}} \, \frac{dx}{\sqrt{F(x) \, G(x)}}.
\label{xproper}
\end{equation}	

As we scan through the parameter space of the AdS C-metrics it is possible to encounter degenerate roots of the function $G(x)$. In that case one encounters a new spatial infinity, for near a double root $x_0$ of $G(x)$ the spatial part of the metric reduces to
\begin{equation}
ds_2^2 = \frac{dx^2}{(x-x_0)^2 } +  (x-x_0)^2 \, d\phi^2,
\label{x0double}
\end{equation}	
which is the metric on a Euclidean hyperboloid $\H^2$. This situation arises when we take $\kappa =1$ and $\mu = \mu_c = \frac{1}{3\,\sqrt{3}}$ and was examined in some detail in \cite{Hubeny:2009hr}. Due to the presence of this internal infinity, geodesics approaching $x_0$ are complete and one has a well defined asymptotic region, viz., $\R \times \H^2$. This was useful in the analysis of \cite{Hubeny:2009hr} since one could disentangle the physics of the black hole horizon from any curved spacetime effects associated with lack of spatial asymptopia. Below, we will investigate all the possible situations that arise from the AdS C-metric in some detail.

%~~~~~~~~~~~~~~~~~~~~~~~~~~~~~~~~~~~~~~~~~~~~~~~
\section{Looking for black funnels and droplets in the AdS C-metric}
\label{s:coordcmet}
%~~~~~~~~~~~~~~~~~~~~~~~~~~~~~~~~~~~~~~~~~~~~~~

We now proceed to identify black funnels and black droplets in the C-metric spacetimes \req{adsc}.  We will find it convenient to separate the discussion into various cases determined by the discrete parameter $\kappa$, and into sub-cases depending on the range we allow for the coordinates $x$ and $y$.  As we will see, for a given C-metric there are in general several different interesting coordinate domains to consider.  In each case, we analyze the situation for all $\mu > 0$ and $\lambda > -1$.  The special limit $\lambda \to -1$ is treated separately in \App{s:lambn1}.

Now, in \cite{Hubeny:2009hr},  droplets and funnels were primarily distinguished by their behavior with respect to the boundary's asymptotic region.  Black droplets had $\gamma$-compact horizons (by which we mean compact with respect to the conformally rescaled metric with the chosen boundary value $\gamma_{\mu\nu}$), and so were well separated from the asymptopia of $\gamma_{\mu\nu}$.  Black droplets were also typically suspended above a second horizon which did not connect to the boundary,  though this was a secondary feature.  In contrast, black funnel horizons were non-compact, and extended into the bulk region associated with the boundary $\gamma_{\mu\nu}$ asymptopia;  see \fig{f:fundrop}.  Specifically,  \cite{Hubeny:2009hr}  required black funnels to asymptote to the bulk black hole solution known to describe a deconfined plasma in the asymptotic region; e.g., a planar AdS black hole for asymptotically flat boundaries or the hyperbolic (aka topological) black hole of \cite{Emparan:1999gf} for boundaries which approach $\mathbb{R} \times \H^2$.

Below, most of our boundary metrics will describe spatially compact universes, with no useful asymptotic regions.  As a result, the definitions of black funnels and droplets given in \cite{Hubeny:2009hr} do not apply and must be generalized.  It is not clear to us what is the right definition in the broadest possible setting, or even whether a sharp distinction between droplets and funnels would remain possible.  However, all the geometries we study below possess a rotational Killing field $\partial_\phi$ which commutes with the static Killing field $\partial_t$.  Furthermore,   each solution contains two special loci defined by $\partial_\phi$.  At least one of these (but possibly both) is an axis (fixed point set) of $\partial_\phi$ corresponding to a root $x_0,x_1,$ or $x_2$ of $G(x)$ where the norm vanishes.  In the one-axis case, the other is the singularity at $x = -\infty$, at which the norm of $\partial_\phi$ diverges.  This structure may be used to give a useful definition as follows.

First, in somewhat of an abuse of language, we refer to any horizon which does not reach the boundary as a {\it planar black hole}, no matter what the geometry or topology.  In our examples below, these will vaguely resemble the planar black hole of \fig{f:fundrop}(b).  It then remains only to classify horizons which connect to the boundary.  Second, we remark that we are interested only in the so-called outer horizons, which by definition are horizons visible from the static region of the boundary.  In some cases, the boundary metric will have two static regions and we will be forced to first choose a particular such region as a reference point.

Now, since the horizons lie at constant $y$ and the axes/singularities of $\partial_\phi$ lie at constant $x$ (and since $x$ and $y$ are independent),
each outer horizon will intersect either one or both of the axes/singularities of $\partial_\phi$.  The case of zero intersections does not arise below.  In addition, it is convenient that each horizon below with two such intersections fails to reach the boundary.\footnote{There is a degenerate case which arises as $\lambda \to -1$ when $y_2 \to x_2$.  In this case, the $y_2$ horizon might be said to develop a second intersection precisely at the boundary.  However, taking the limit $\lambda \to -1$ carefully so as to maintain the AdS radius $\ell_4$  fixed leads to a solution with only a single intersection for each horizon which reaches the boundary. See \App{s:lambn1}.}  As such, these are planar black holes in the sense described above and we need not consider this case further.  It remains to classify the cases with a single intersection, which we do as follows:

\begin{itemize}
\item{\bf Black funnel:}  When following a $\partial_\phi$ axis or singularity {\it outward} from the horizon leads toward the boundary,  we call the horizon a black funnel.  Unless a second horizon is encountered, the axis or singularity then connects the horizon to the boundary through the visible static region. 
In such cases, an artistic impression of the spacetime resembles \fig{f:fundrop}(a), with the axis or singularity located near the edge of the diagram.  Below, this is always an axis as the singularity at $x= -\infty$ is always hidden by a horizon on the boundary.   In some sense, the point where the axis reaches the boundary plays the role of an asymptotic region for the boundary metric.  When black funnels appear below, only one axis or singularity of $\partial_\phi$ will be visible from the static region of the boundary.

\item {\bf Black droplet:}  When following a $\partial_\phi$ axis or singularity inward from the horizon leads toward the boundary,  we call the horizon a black droplet.  Unless a second horizon is encountered, the axis or singularity then connects the horizon to the boundary through a hidden region behind the horizon.    In the cases that arise below, we will always find a second separate $\partial_\phi$ axis or singularity that intersects the boundary outside the droplet horizon. One may think that the first axis or singularity plays the role of the origin while the second axis or singularity plays the role of an asymptotic region in the boundary metric.  In this sense, an artistic impression of the spacetime resembles \fig{f:fundrop}(b).  In cases that appear below, the first axis or singularity also extends ``below'' the droplet to intersect a planar black hole, a black funnel, or a new singularity.
\end{itemize}

As the reader will note, the above definitions are based on geometric features of horizons which generalize those of funnels, droplets, and planar black holes as defined in \cite{Hubeny:2009hr} and illustrated in \fig{f:fundrop}.  We nevertheless conjecture that, as for the original definition in \cite{Hubeny:2009hr}, the funnel solutions are dual to field theory black holes coupling strongly to a deconfined plasma, while the droplets are dual to black holes coupling weakly.  Some evidence for this is provided by the fact that all droplet solutions below which are free of naked singularities also include a second disconnected outer horizon.  This second horizon can be interpreted as describing the plasma, and the lack of connection as a sign of the weak coupling.  See \sec{s:cbdystress} for further discussion.

%~~~~~~~~~~~~~~~~~~~~~~~~~~~~~~~~~~~~~~~~~~~~~~~
\subsection{Case A: Funnels and  droplets for $\kappa = 1$}
\label{s:k1fd}
%~~~~~~~~~~~~~~~~~~~~~~~~~~~~~~~~~~~~~~~~~~~~~~

We begin our discussion with  the case $\kappa =1$.  For $\mu < \mu_c \equiv\frac{1}{3\,\sqrt{3}}$ we find that $G(x)$ has three real roots and for $\mu > \mu_c$ we obtain a single real root. Precisely for $\mu = \mu_c$ we have a degenerate root of $G(x)$ (this condition determines $\mu_c$) --  this was the situation discussed in detail in \cite{Hubeny:2009hr}. As $\mu \to \mu_c$ from below, the roots $x_0$ and $x_1$ approach each other.  For larger values of $\mu$ these roots move off into the complex plane.

Now that we have understood the roots of $G$, we can answer an important basic question: What restriction should we impose on the ranges of the coordinates $x$ and $y$? In order to maintain the correct Lorentzian signature of the metric, we require $G(x) \ge 0$.  For $\mu < \mu_c$, this means that to avoid unwanted boundaries at finite distance we should consider either the region $x_1 \le x \le x_2$ or alternately $x \le x_0$. The former is the conventional choice that has been made in previous analyses of the AdS C-metric \cite{Emparan:1999wa,Emparan:1999fd,Dias:2002mi}, but  the latter is a perfectly reasonable coordinate domain as well.  For  $\mu > \mu_c$ we must take $x \le x_2$.  We should also determine whether we restrict attention to the region $y\le x$ or $y\ge x$.  To this end it is useful to introduce a new coordinate
\begin{equation}
z = x- y,
\label{coord1}
\end{equation}	
which keeps track of the distance from the boundary $x=y$, so that we choose either $z <0$ or $z > 0$. It is worth noting that the coordinate $z$ introduced in \req{coord1} is not the conventional Feffereman-Graham coordinate, which would instead be defined via the gauge choice $g_{zz} = z^{-2}$ and $g_{z\mu} =0$.
% Figure

As a final piece of preparation, we describe the root structure of $F(y)$, which will in turn determine the horizons.  Recalling that the roots of $F$ and $G$ are ordered by \req{order} (with the exceptions noted in \sec{s:geocmet}),
it is easy to convince oneself that $F(y)$ behaves as follows for $\kappa=1$:
\begin{list2}
\item[1.] For $\lambda >0$ we have $F(\xi) >0$ for $\xi >0$.  Thus from \req{order} we have a single real root at $y_0 < 0$.
\item[2.] For $\lambda =0$ we encounter a double root at the origin, which is degenerate and corresponds to the bulk Poincar\'e horizon.
\item[3.] For  $\lambda \in (\text{max}\{-\frac{1}{27\,\mu^2} ,-1\},0)$ we have three real roots of the function $F(\xi)$, two of which are negative.
\item[4.]  In the special case when $\lambda =  -\frac{1}{27\,\mu^2}$ (and $\mu \ge \mu_c$), we have a double root at $y_0 = y_1  = -\frac{1}{3\,\mu}$ in addition to a single root at $y_2 = \frac{1}{6\,\mu}$.
\item[5.] For $-1< \lambda <-\frac{1}{27\,\mu^2}$ there is only one real root occurring at some $y_0 >0$. Of course, this domain is non-empty only for $\mu > \mu_c$.
\end{list2}
Note that, due to our convention that the root be called $y_0$ whenever $F$ has only one real root, the function $y_0(\lambda)$ is discontinuous at $\lambda = -\frac{1}{27\,\mu^2}$.  The single root for $\lambda < -\frac{1}{27\,\mu^2}$ actually  continuously connects to $y_2(\lambda)$ for $\lambda \ge -\frac{1}{27\,\mu^2}$.  This is the cause of the main exception to (\ref{order}) noted in \sec{s:geocmet}.

\begin{figure}[hp]
\begin{center}
\includegraphics[scale=0.8]{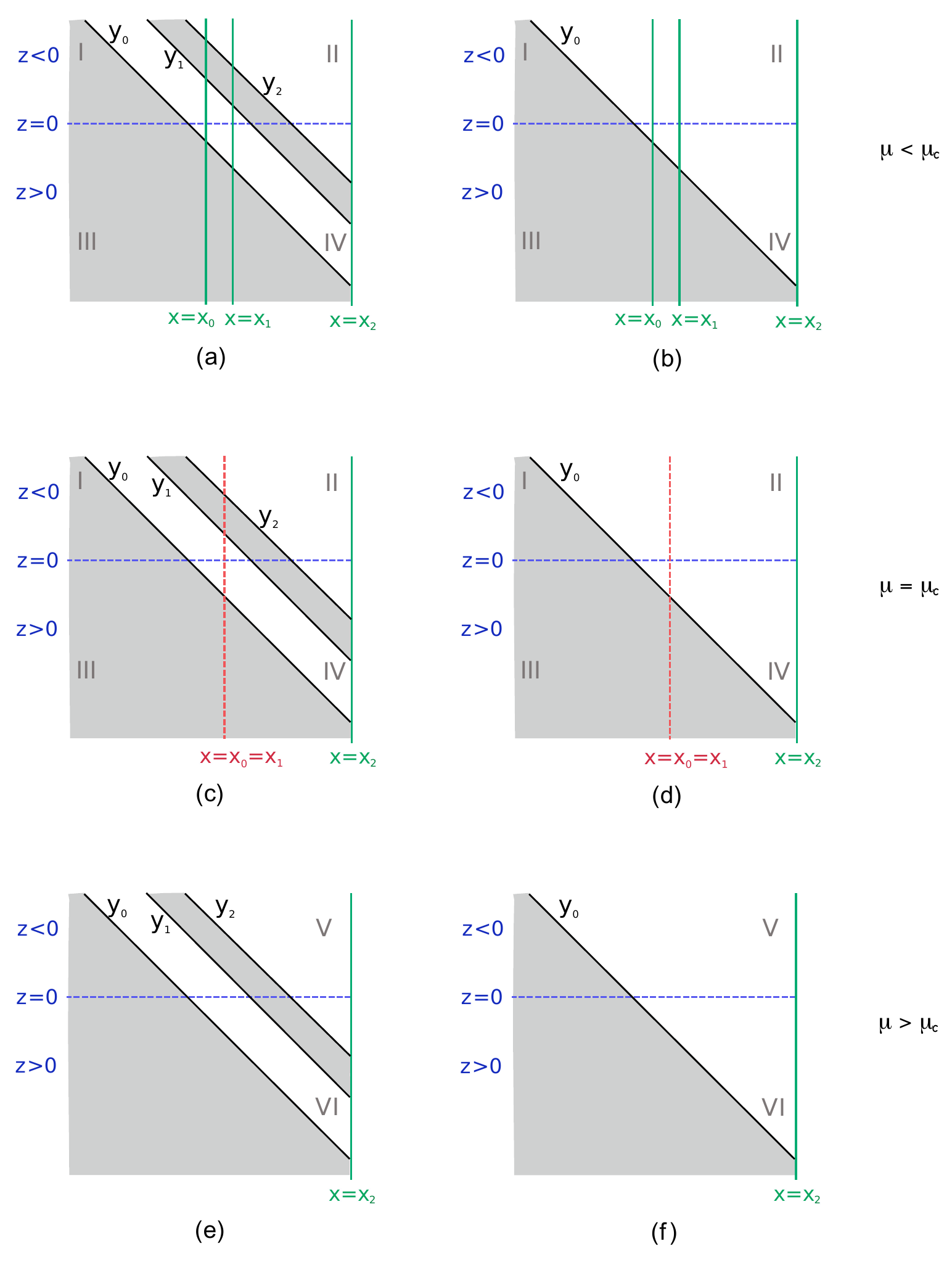} \\
\end{center}
\caption{A sketch of the possible coordinate domains for the AdS C-metric with $\kappa =1$ for various values of $\mu$.  Horizons (diagonal lines) are plotted in the $(x,z)$ plane. Note that $z$ increases downward while $x$ increases to the right. The allowed regions are indicated by the roman numerals and can be considered as a complete spacetime unto themselves. To maintain the correct Lorentz signature, the allowed regions are $x \le x_0$ and $x \in [x_1,x_2]$ respectively which are indicated by the numbers. The different panels for a given value of $\mu$ correspond to situations with different numbers of roots for $F(x)$; for a detailed behavior of the roots see \fig{f:fgplots}.}	
\label{f:regionk1}
\end{figure}

We are now in a position to list the various possible coordinate domains and to analyze each in turn as a function of $\lambda$ and $\mu$.  For a complete illustration of the horizons in each domain, see \fig{f:regionk1}.

% Figure

\noindent
{\bf Region A-I: }  $x \in (-\infty, x_0]\; \& \;z \le0$: This situation occurs for $\mu \le \mu_c$ (panels (a), (b), (c), and (d) of \fig{f:regionk1}).  On the boundary, we encounter a  horizon at $x=y_0$. This boundary horizon is akin to the cosmological horizon in deSitter space, with the $\phi$-circle pinching off as one moves outward to $x=x_0$.  However, from the boundary perspective,  the horizon shields a curvature singularity at $x = -\infty$.

In contrast, the $x= -\infty$ singularity is visible through the bulk and in fact intersects the bulk $y=y_0$ horizon.  Following the singularity away from $y_0$ through the static region takes us away from the boundary, so $y_0$ describes a (singular) black droplet.  The droplet does not reach the axis at $x_0$, and this axis is visible from the boundary.  However, for all $\mu < \mu_c$ we can pick the period of the angle $\phi$ to be $\Delta \phi = \frac{4\pi}{|G'(x_0)|}$ to avoid a conical singularity.

In the domain $\lambda < 0$ where $F(y)$ has three real roots (panels (a)  and (c) of \fig{f:regionk1}), the $y_0$ droplet is suspended above a (singular) planar black hole horizon at $y= y_1$, i.e., we have two disconnected outer horizons. Note that there is a third horizon at $y=y_2$ which also reaches to $x = x_0$.  However, this is an inner horizon since it is always hidden from the boundary observer.

The special case $\mu =\mu_c$ (panels (c) and (d) of \fig{f:regionk1})
where we encounter a double root at $x_0$ ($x_1$ coalesces with $x_0$ in this limit) was studied in some detail in \cite{Hubeny:2009hr}. In this case the spatial metric on the boundary is locally $\H^2$ near $x=x_0$ and one has a well behaved asymptotic region of the boundary metric.  Since the axis has moved off to infinity, conical singularities do not arise for any choice of $\Delta \phi$, though the curvature singularity at $x= -\infty$ remains visible.   For $\lambda < 0$ (panel (c)) we have a (singular) black droplet suspended over a (singular) hyperbolic AdS black hole in the bulk, though the latter horizon disappears for $\lambda >0$ (panel (d)). 

\noindent
{\bf Region A-II: }  $x \in [x_1, x_2]\; \& \; z\le 0$. This situation similarly arises when $\mu \le \mu_c$ (panels (a) - (d)).
Since the $\phi$-circle shrinks as one moves away from the horizon, any static region of the boundary spacetime again vaguely resembles that of de Sitter space.

The case $\lambda > 0$ (panels (b) and (d)) is uninteresting, as it contains no horizons in either the bulk or boundary.  It has only a naked singularity at
$y = +\infty$.
For $\lambda <0$
and $\mu < \mu_c$ (panel (a)),
 the cubic $F(y)$ has two roots which satisfy $x_1 < y_1 < y_2 < x_2$, indicating that there are two black hole horizons  on the boundary.  This also leaves us with two static regions on the boundary: $y_2 < x < x_2$, or $x_1 < x < y_1$.  Note that for a given choice of static region, only one horizon will be an outer horizon. The singularity at $y = +\infty$ is visible from the former static region, so we focus on the latter.  From this perspective, $y_1$ is an outer horizon and $y_2$ an inner horizon. Both horizons reach the axis at $x = x_1$, so $y_1$ is a black funnel.  (Curiously, the inner horizon of the black funnel would in fact look like a droplet from the other static region's perspective.)
 The $x_2$ axis is hidden behind the horizon, so we avoid all naked singularities by taking $\Delta \phi = \frac{4\pi}{|G'(x_1)|}$.  The $y_1$ and $y_2$ horizons merge to form a smooth extreme horizon when $\lambda =0$.

The spacetime is similar in the limit  $\mu \to \mu_c$ (panel (c)),  though the $x_0$ and $x_1$ axes merge and move off to infinite distance, creating a new asymptotic region near $x= x_0 = x_1$; see \cite{Hubeny:2009hr} for details.     As a result, for $\lambda < 0$ no conical singularities are visible from the static region $x_1 < x < y_1$ for any value of $\Delta \phi$.

\noindent
{\bf Region A-III: } $x \in (-\infty, x_0]\; \& \; z \ge0$. We again encounter this situation only for $\mu \le \mu_c$ (i.e.\ panels (a) - (d)).  We have a black hole on the boundary with $x = y_0$,
and we have a bulk horizon which starts from $x=y_0$ on the boundary and reaches the axis $x=x_0$ in the bulk.  This axis connects the horizon and boundary through a static region, so the $y_0$ horizon is a black funnel.   There are no visible curvature singularities, and the period of $\phi$ may be chosen to make the axis at $x_0$ regular. The situation is similar when $\mu = \mu_c$, though since $x_0$ now represents an asymptotic region there is no conical singularity for any value of $\Delta \phi$.

\noindent
{\bf Region A-IV: } $x \in [x_1, x_2]\; \& \; z \ge 0$: The regime of parameter space where we encounter this possibility is as in {\bf A-II} given by $\mu \le \mu_c$. The only difference is that we allow ourselves to consider a different range of the coordinate $z$.

Consider first $\mu < \mu_c$.
For $\lambda > 0$ (panel (b)) there is only a planar black hole; the boundary metric contains no horizon.
However, there are boundary black holes for $\lambda \le  0$ (panel (a)). In fact, as in {\bf A-II},
the boundary has two static regions: $y_2 < x < x_2$, or $x_1 < x < y_1$.  From the perspective of the former, only $y_2$ is an outer horizon.  It forms a black funnel.  Since only the $x_2$ axis is visible, the choice $\Delta \phi = \frac{4\pi}{|G'(x_2)|}$ leaves no naked singularities.  From the perspective of the latter region, the outer horizons are $y_0$ (a planar black hole) and $y_1$ (a droplet).  Since both axes are visible there is a naked conical singularity for any choice of $\Delta \phi$, though all curvature singularities are hidden.  The $y_1$ and $y_2$ horizons merge to form a smooth extreme horizon as $\lambda \to 0$.

The situation is similar for $\mu =\mu_c$. For $\lambda > 0$ (panel (d))  there is only a planar black hole, but for $\lambda \le 0$ (panel (c)) there are again two static regions of the boundary metric: $y_2 < x < x_2$, or $x_1 < x < y_1$.    In fact, from the perspective of the first static region, the situation is identical to that with $\mu < \mu_c$.  The new asymptotic region at $x_0$ is visible only from the latter static region, where it replaces an axis and allows us to remove all conical singularities by choosing $\Delta \phi = \frac{4\pi}{|G'(x_2)|}$, so that no singularities are visible.   We have a black funnel at $y_2$ and a black droplet at $y_1$, suspended above the planar horizon at $y_0$.  As described in \cite{Hubeny:2009hr}, this planar horizon asymptotes near $x_0$ to one of the hyperbolic black holes described in \cite{Emparan:1999gf}.

\noindent
{\bf Region A-V: } $x \in (-\infty, x_2]\; \& $ $z \le 0$:
We encounter this possibility when $\mu > \mu_c$ since in that regime $G(x)$ has a single root at some $x =x_2> 0$.  From (\ref{order}), we see that the boundary always contains a black hole.  When $F$ has only one real root (panel (f)), it describes a single bulk horizon at $y_0$ which intersects the singularity at $x = -\infty$.  It is a singular droplet.  The singularity at $y = +\infty$ is also visible, as is the $x_2$ axis.    For $-1< \lambda <-\frac{1}{27\,\mu^2}$, $F$ has three roots (panel (e)) and there are two choices of static region on the boundary, $y_0 < x < y_1$ and $y_2 < x < x_2$.  From the perspective of the first, the $y_0$ horizon is a singular droplet suspended above a (singular) funnel at $y_1$.
From the perspective of the second, the $y_2$ horizon is again a singular droplet and, in addition, the $x_2$ axis is visible.

\noindent
{\bf Region A-VI: } $x \in (-\infty, x_2]\; \& \; z \ge 0$:
As in the preceding case, this pertains to $\mu > \mu_c$.
When $F$ has a single real root (panel (f)),  the only horizon is $y_0$.  It is a black funnel and the $x_2$ axis is visible.  For $-1< \lambda <-\frac{1}{27\,\mu^2}$, $F$ has three real roots (panel (e))  and we again find two static regions on the boundary.    Choosing $y_0 < x < y_1$, the $y_1$ horizon is a black droplet suspended above a black funnel at $y_0$.  Choosing $y_2 < x < x_2$, the $y_2$ horizon is a black funnel.   In both cases, setting  $\Delta \phi = \frac{4\pi}{|G'(x_2)|}$ is necessary and sufficient to avoid naked singularities.

%~~~~~~~~~~~~~~~~~~~~~~~~~~~~~~~~~~~~~~~~~~~~~~~
\subsection{Case B: Funnels and  droplets for $\kappa = 0$}
\label{s:k0fd}
%~~~~~~~~~~~~~~~~~~~~~~~~~~~~~~~~~~~~~~~~~~~~~~

We next consider the simple case $\kappa =0$, where it is clear that the only real root of $ G(x)$ is located at $x_2 = \left(\frac{1}{2\,\mu}\right)^{\! \!  1/3}$.  Likewise $F(y)$ has a single real root at $y= y_0 =\left(- \frac{\lambda}{2\,\mu}\right)^{\! \! 1/3}$, so this is the only horizon. This root is non-degenerate for $\lambda \neq 0$, but becomes triply degenerate at $\lambda =0$.  From \req{order} we see that the boundary metric has  a single static region $y_0 < x < x_2$, from which the $x_2$ axis is clearly visible.  However, $y_0$ can lie on either side of the origin, depending on the sign of $\lambda$.

Maintaining Lorentz signature of the metric requires that we restrict attention to $x \le x_2$.  The only choice for the coordinate range is whether we approach the boundary at $x=y$ from above or below; both choices yielding the same boundary metric.  Moreover, since there is only one root of the function $G(x)$, we can choose $\phi$ to have the correct period to get rid of the potential conical defect, i.e., $\Delta \phi = \frac{4\pi}{|G'(x_2)|}$.  Once again, the situation is analogous to that encountered in  the static region of de Sitter space:  the spatial sections are compact, and the size of the $\phi$ circle decreases as one moves away from the horizon, shrinking to zero at the $x_2$-axis.

% Figure
\begin{figure}[tp]
\begin{center}
\includegraphics[scale=1]{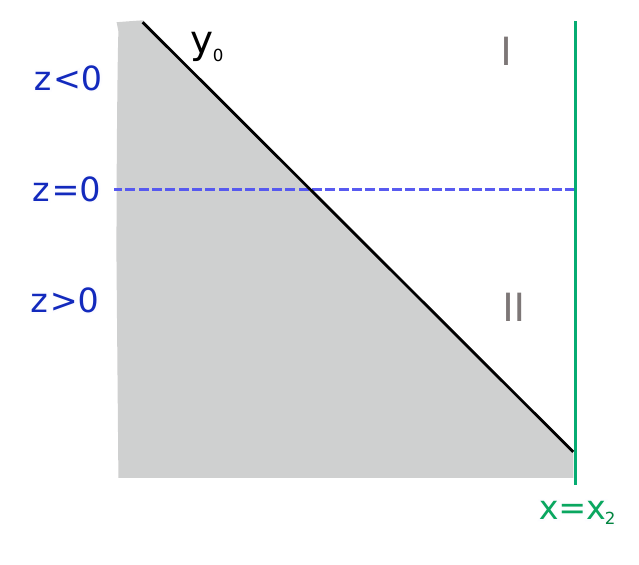}
  \caption{A sketch of the possible coordinate domains for the AdS C-metric with $\kappa =0$.    Same conventions are used here as in \fig{f:regionk1}.}	
\label{f:regionk0}
\end{center}
\end{figure}

The various possible coordinate regions are as shown in \fig{f:regionk0} and can be summarized as follows.

\noindent
{\bf Region B-I:} $x \le x_2 \; \& $ $z \le 0$.
We have a black hole horizon on the boundary
which extends away towards large negative $x$ in the bulk, reaching the singularity at $x = -\infty$. Following the singularity away from the horizon through the static region one moves away from the boundary, so the horizon is a (singular) black droplet.  The $x_2$ axis and the singularity at $y = +\infty$ are visible from the boundary.

\noindent
{\bf Region B-II:} $x \le x_2 \; \& $ $z \ge 0$.
The horizon is a black funnel.  For $\Delta \phi = \frac{4\pi}{|G'(x_2)|}$, the only singularities occur at $x = -\infty$ and $y = -\infty$.  Both singularities are hidden behind the horizon.

%~~~~~~~~~~~~~~~~~~~~~~~~~~~~~~~~~~~~~~~~~~~~~~~
\subsection{Case C: Funnels and  droplets for $\kappa = -1$}
\label{s:km1fd}
%~~~~~~~~~~~~~~~~~~~~~~~~~~~~~~~~~~~~~~~~~~~~~~

 Finally, for $\kappa = -1$ one has only a single real root for $G(x)$ at some $x_2 >0$. This is clear from \req{order} and the fact that now $G(x) > 0$ for all $x \le 0$. We must therefore allow the entire range $x \le x_2$.   % $\kappa =0,1$, the only choice is 
 \bit{However, as for $\kappa =0,1$, one must choose} whether to take $z\ge0$ or $z\le 0$.

% Figure
\begin{figure}[tp]
\begin{center}
\includegraphics[scale=0.9]{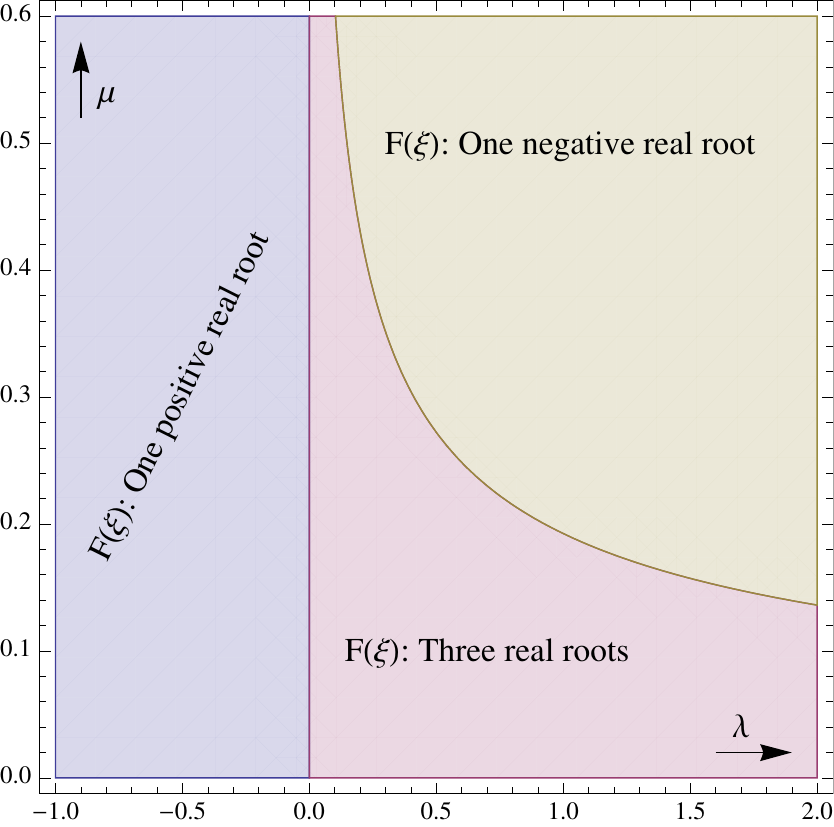}
  \caption{A plot of the domains in the $\{\lambda, \mu\}$ plane
%  where the functions $F(\xi)$ and $G(\xi)$ have different properties for their real roots
    which characterize the distinct possibilities for the root structure of $F(\xi)$
  for $\kappa =-1$. Note that the behavior of $G(\xi)$ is universal; it always has one positive real root.}	
\label{f:kappam1dom}
\end{center}
\end{figure}

Once again we can analyze the behavior of $F(y)$ and use \req{order} to conclude that (see \fig{f:kappam1dom} for an illustration):
\begin{list2}
\item[1.]  For $-1< \lambda < 0$ we have a single real root for some $0<y_0 <x_2$.
\item[2.] For $\lambda = 0$ there is a degenerate root at the origin and a positive real root at $y_2 < x_0$.
\item[3.] For $\lambda \in (0, \frac{1}{27\, \mu^2})$ there are three real roots, one of which is negative and the other two positive, which we order as $y_0 < 0 < y_1 < y_2 < x_2$.

\item[4.] For $\lambda = \frac{1}{27 \mu^2}$, $F$ has a single root at  $-1/6\mu$ and a double root at  $1/3\mu$.
\item[5.] For $\lambda > \frac{1}{27\, \mu^2}$ we have a single negative real root at $y= y_0$ for $F(y)$.
\end{list2}
%

% Figure
\begin{figure}[tp]
\begin{center}
\includegraphics[scale=1]{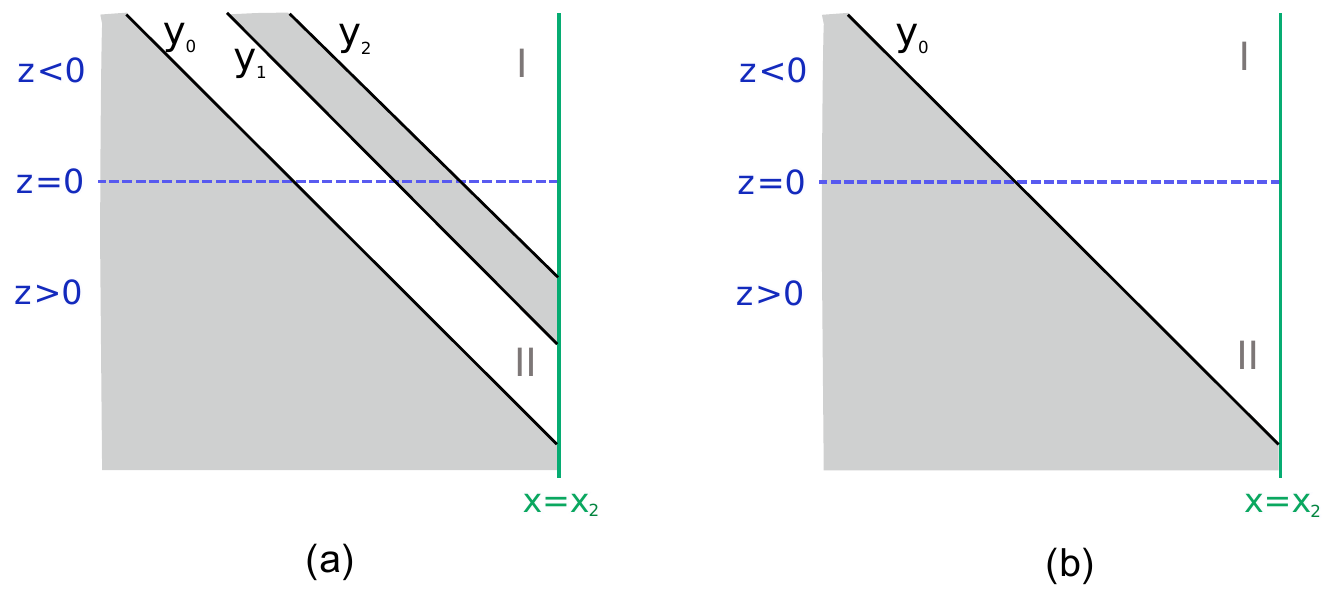}
  \caption{A sketch of the possible coordinate domains for the AdS C-metric with $\kappa =-1$.
  Same conventions are used here as in \fig{f:regionk1}.
The allowed regions are $x \le x_2$. The left panel corresponds to the situation when $F(y)$ has three real roots,  while the right panel protrays the situations when it has just one, compare with \fig{f:kappam1dom}.}
\label{f:regionk1mgmc}
\end{center}
\end{figure}
Note that, due to our convention that the root be called $y_0$ whenever $F$ has only one real root, the function $y_0(\lambda)$ is discontinuous at $\lambda =0$.  The single root for $\lambda < 0$ actually  continuously connects to $y_2(\lambda)$ for $\lambda \ge 0$.

We therefore find the following behaviors:

\noindent
{\bf Region C-I:}  $x \le x_2$ \& $z \le 0$.   The singularity at $x = -\infty$ is always visible through the bulk.  When $F$ has only one real root (panel (b) of \fig{f:regionk1mgmc}), we find a singular  droplet.  The $x_2$ axis and the $y = +\infty$ singularity are also visible.   When $F$ has three real roots (panel (a)), we have a choice of static regions on the boundary.  From the perspective of the region
$y_0 < x < y_1$, the $y_0$ horizon is a singular droplet suspended above a (singular) funnel at  $y_1$.  However, the $y = +\infty$ singularity and the $x_2$ axis are hidden.  From the perspective of the region
 $y_2 < x < x_2$, the $y_2$ horizon is a singular droplet and the $x_2$ axis and $y = +\infty$ singularity are visible.

\noindent
{\bf Region C-II:}  $x \le x_2$ \& $z \ge 0$.   When $F$ has only one real root,  there is a single horizon at $y_0$.  It is a black funnel.  When $F$ has three real roots,  we have a choice of static regions. From the perspective of the region
$y_0 < x < y_1$, the $y_1$ horizon is a droplet suspended above a black funnel at $y_0$.   From the perspective of the region
 $y_2 < x < x_2$, the $y_2$ horizon is a black funnel.
In all cases, the choice  $\Delta \phi = \frac{4\pi}{|G'(x_2)|}$ is necessary and sufficient to avoid naked singularities.

%~~~~~~~~~~~~~~~~~~~~~~~~~~~~~~~~~~~~~~~~~~~~~~~
\section{Boundary stress tensor for the C-metric}
\label{s:cbdystress}
%~~~~~~~~~~~~~~~~~~~~~~~~~~~~~~~~~~~~~~~~~~~~~~~~

We conclude by computing the boundary stress tensor for the solutions described above. This will allow us to see whether we can interpret any region around the back hole as containing a thermal fluid, in which one might hope to in some sense separate the effects of Hawking radiation from those of vacuum polarization.   To this end, we need to use either (i) an explicit coordinate transformation to Fefferman-Graham coordinates or (ii) an appropriate definition of the boundary and use the counter-term procedure to compute the stress tensor. We will follow the latter strategy since it is more convenient to implement for our purposes.

The counter-term procedure outlined in \cite{Balasubramanian:1999re}  adds a boundary term to the Einstein-Hilbert action so that the result provides a well-defined variational principle for asymptotically AdS spacetimes \cite{Papadimitriou:2005ii}.  The full action is 
\begin{equation}
\CS = \frac{1}{16\pi\,G_N^{(4)}}\, \int \, d^4x\, \sqrt{-g}\, \left(R - 2\, \Lambda_4\right) + \frac{1}{16\pi\,G_N^{(4)}}\,  \int\, d^3x \, \sqrt{-\gamma} \left( 2\, K  - \frac{2}{\ell_4} + \frac{\ell_4}{2}\, \CR\right).
\label{EHAction}
\end{equation}	
Here $\gamma$ is the induced metric on the boundary, which we take to be a surface of constant $z = x-y$.  Similarly, $K$ is the extrinsic curvature, and $\CR$ the boundary Ricci scalar. It is important to note that this $z$ agrees with the Fefferman-Graham coordinate typically used in the holographic renormalization literature only to leading order, and not beyond.

Variations of \req{EHAction} with respect to $\gamma^{\mu\nu}$ lead to the stress tensor-like object
\begin{equation}
16\pi\,G_N^{(4)}\, \CT_{\mu \nu} = K_{\mu \nu} - \gamma_{\mu \nu}\, K - \frac{2}{\ell_4} \, \gamma_{\mu \nu} + \ell_4\, \left(\CR_{\mu\nu} - \frac{1}{2}\, \CR\, \gamma_{\mu\nu}\right).
\label{sten1}
\end{equation}	
However, because the metric $\gamma_{\mu \nu}$ diverges on the boundary, some rescaling will be required to obtain the  boundary stress tensor.  One notes that $\gamma^{\mu \nu}  = e^{-2\phi} \, \tilde{\gamma}^{\mu \nu}$, where $e^{\phi} = z/\ell$ and $\tilde \gamma^{\mu \nu}$ is the inverse of the physical boundary metric \req{indbdymet}.

Since $T^{\mu\nu}$ has conformal dimension five, we have $T^{\mu \nu} = e^{-5 \,\phi} \,\tilde{T}^{\mu \nu}$, where $\tilde{T}^{\mu \nu}$ is the physical stress tensor on the boundary.  As a result,  the stress tensor with lower components will have a single factor of $e^{-\phi}$:
\begin{equation}
T_{\mu\nu} = \lim_{z \to 0} \,\frac{\ell}{z}\, \CT_{\mu\nu}.
\label{bdystress}
\end{equation}	

For the boundary metric \req{indbdymet} we obtain
\begin{eqnarray}
T_{tt} &=&   \frac{c\, \mu}{\sqrt{1+\lambda}} \, \gamma_{tt}\,  \left[G(x)  -2\, F(x)\right] \nonumber \\
T_{xx} &=&  c\, \mu\,\sqrt{1+\lambda} \, \gamma_{xx}\,
\nonumber \\
T_{\phi\phi} &=&  c\, \frac{ \mu}{\sqrt{1+\lambda}} \,  \gamma_{\phi\phi} \, \left[F(x) - 2 \, G(x) \right],
\label{sten}
\end{eqnarray}	
where we define a central charge $c\equiv \frac{\ell_4^2}{16\pi\,G_N^{(4)}}$, measuring the effective degrees of freedom.\footnote{E.g., for \AdS{4} geometries obtained by compactifying M-theory on Sasaki-Einstein seven-folds we have $c \propto N^{3/2}$ where $N$ indicates the number of M2-branes probing the singularity of the Calabi-Yau cone over the Sasaki-Einstein base.}  We have also simplified the expression using  the fact that   $F'''(x) = 12\, \mu$. In deriving this expression we used the relation \req{ads4cscale} to express the parameter $\ell$ in terms of the physical length scale $\ell_4$. Note that the signs and factors of two flip between the $tt$ and $\phi\phi$ components as we pass from horizon to the axes, reflecting the symmetry between $x$ and $y$ in the Euclidean-signature solution.
Finally, since the boundary field theory is odd dimensional, there is no conformal anomaly and the stress tensor must be traceless. Happily, tracelessness follows from the relation $F(x) + G(x) = 1+\lambda$.  A more natural way to  present our result which makes the traceleness obvious is given by
\begin{equation}
T^\mu_\nu = c\, \frac{ \mu}{\sqrt{1+\lambda}} \,  \; \text{diag}\bigg\{ G(x) - 2 \, F(x) , F(x) + G(x) , F(x) - 2\, G(x)\bigg\}.
\label{Tupdn}
\end{equation}	

In general, \req{Tupdn} does not take the perfect fluid form
\begin{equation}
{\bf T}_{\mu\nu} = P(x) \,\left(3\, u_\mu \, u_\nu + \gamma_{\mu \nu}\right)
\label{thermst}
\end{equation}	
due to vacuum polarization effects.  However, we note that \req{Tupdn} does reduce to \req{thermst} at any root of $G(x)$ where $\partial_t$ is timelike ($F(x) > 0$); i.e., where an axis of $\partial_\phi$ intersects the static region of the boundary.  There we identify $u^\mu =  \frac{1}{\sqrt{\gamma_{tt}}}\, \left(\frac{\p}{\p t}\right)^\mu$ and $P(x) = c \, \frac{\mu}{\sqrt{1+\lambda}} \, F(x) = c \,\mu \,\sqrt{1+\lambda}$.   This is an analogue of the fact that for
$\mu = \mu_c$  and $\kappa=1$ this occurs in the asymptotic region near $x=x_0$,  where we saw in \cite{Hubeny:2009hr} that the C-metric solutions approach the hyperbolic black holes of \cite{Emparan:1999gf} and are dual to a thermal plasma.

Note that for our funnel solutions any axis of the above type intersects the funnel horizon, while for droplet solutions such an axis cannot intersect the droplet horizon.  Instead, in every droplet case it intersects either a naked singularity, a planar black hole, or a black funnel.  Though there is no sharp argument, we take this as supporting our basic picture of funnels and planar black holes as describing plasmas near the above axes, while droplet horizons describe physics that is only weakly coupled to the plasma.

%~~~~~~~~~~~~~~~~~~~~~~~~~~~~~~~~~~~~~~~~~~~~~~
\section{Discussion}
\label{s:discuss}
%~~~~~~~~~~~~~~~~~~~~~~~~~~~~~~~~~~~~~~~~~~~~~~

The AdS C-metric has been the inspiration for many interesting solutions in the past, most notably the brane-world black holes constructed in \cite{Emparan:1999wa, Emparan:1999fd} and more recently the exact plasma ball solution of \cite{Emparan:2009dj}.
We have in this paper undertaken a detailed analysis of the static, uncharged AdS C-metrics to infer the examples of black funnels and black droplet solutions hidden in this family.   This required a generalization of the definitions of funnel and droplet given in  \cite{Hubeny:2009hr} to boundary metrics describing spatially compact universes.

As in \cite{Hubeny:2009hr}, we conjecture that such solutions correspond to states of strongly coupled CFTs in black hole backgrounds, with the distinction of funnel vs.\ droplet corresponding to two different types of behavior for the field theory state. Black funnels appear to describe horizons in the boundary metric coupling strongly to field theory plasmas, while black droplets appear to describe weak such couplings.   In the latter case, the weak coupling is signified by the fact that, for the cases without naked singularities, our droplets were always accompanied by a second disconnected horizon (which we called a planar black hole).  From the field theory perspective we interpret the droplet itself as describing vacuum polarization around the horizon in the boundary metric, and we interpret the planar black hole as describing the plasma.  The fact that these horizons do not meet in the bulk implies that they describe pieces of field theory physics that can be thought of as coupling very weakly in the large $N$ limit.  This is what allows such states to be stationary even though, as one may check, the two horizons always have different temperatures outside of the special $\lambda \to -1$ limits described in \App{s:lambn1}.

The AdS C-metric family contains a rich variety of such solutions.
 The comprehensive set of all possibilities is summarized in \fig{f:regionk1}, \fig{f:regionk0}, and \fig{f:regionk1mgmc} (for $\kappa = 1,0,-1$, respectively), where distinct panels portray distinct ranges of $\mu$ and $\lambda$ yielding qualitatively different behavior, while within individual panels distinct spacetimes are separated by dashed lines (and labeled by Roman numerals).
Indeed, with the definitions given in \sec{s:coordcmet}, for $\mu \neq 0$ every outer horizon could be classified as a planar black hole, a black droplet, or a black funnel.  For each choice of C-metric parameters, we have also identified the static regions from which all singularities are hidden behind horizons.  In these cases, the dual field theory states should be regular at least on the region of the boundary that lies in the given static region.  We also found many settings with naked singularities in the bulk, but for which both the boundary metric and the boundary stress tensor are smooth.  While such solutions are likely to be dual to singular states of the conformal field theory, it would be interesting to understand whether such singularities could be resolved by stringy or quantum effects in the bulk or, more likely, by introducing either time-dependence or some deformation of the dual field theory; e.g., by some perturbation that causes the theory to confine at an energy scale high enough to hide the would-be bulk singularities.

As we have seen, generic values of the C-metric parameters describe boundary metrics which are spatially compact in a natural conformal frame.  One of the disadvantages of this feature is that in the absence of spatial asymptopia it is a-priori unclear how one can disentangle the physics of the black hole horizon from curved spacetime effects. Nevertheless, it is curious that the quasi-local stress tensor induced on the boundary reduces to a thermal perfect fluid form close to any axis (fixed point locus of the spatial isometry $\p_\phi$), at least for certain choices of parameters.   Of course, for special values of parameters it is possible to ensure that the induced boundary metric has a well defined asymptotics, viz., a hyperbolic cylinder $\R \times \H^2$ as studied in \cite{Hubeny:2009hr}.  One may also introduce a new asymptotic region at any point $x_+$ by a change of conformal frame, though at least one of $||\partial_t ||$ or $|| \partial_\phi ||$ will then diverge at $x_+$.  As a result, while this may be useful for studying asymptotically AdS black holes on the boundary, it will not provide boundary black holes with other familiar asymptotic behaviors.

One of the interesting generalizations which we have not explored here is the case of rotating AdS C-metrics. An asymptotically flat rotating black hole spacetime does not have a Hartle-Hawking vacuum due to super-radiance effects \cite{Kay:1988mu}.  For very similar reasons, it is hard to imagine rigidly rotating black funnels when the boundary has a well-defined asymptotic region. This would seem to require the distant shoulders of the black funnel to in some sense rotate faster than the speed of light.  However, the situation is somewhat different in the spatially compact case or when the boundary metric itself is asymptotically AdS.  The more general AdS C-metrics found in \cite{Plebanski:1976gy} do allow for rotation and it would be interesting to examine this issue in some detail.   The related case of rotating BTZ boundary metrics will be studied in  \cite{toappear}.

%_____________________________________________
\subsection*{Acknowledgements}
%_________________________________________________

It is a pleasure to thank Roberto Emparan and  Rob Myers for very interesting discussions.  VEH and MR would like to thank the KITP for wonderful hospitality during the workshop ``Fundamental Aspects of Superstring Theory", as well as the  Pedro Pascual Benasque Center of Science and the Aspen Center for Physics for excellent hospitality during the course of this project. In addition, VEH, DM and MR would like to thank the ICTS, TIFR for hospitality during the Monsoon workshop in string theory where this project was initiated. VEH and MR are supported in part by STFC Rolling grant and by the US National Science Foundation under the Grant No.\ NSF PHY05-51164.  DM was supported in part by the US National Science Foundation under grants   PHY05-55669 and PHY08-55415 and by funds from the University of California.

\appendix

%~~~~~~~~~~~~~~~~~~~~~~~~~~~~~~~~~~~~~~~~~~~~~~~
\section{The limit $\lambda \to -1$}
\label{s:lambn1}
%~~~~~~~~~~~~~~~~~~~~~~~~~~~~~~~~~~~~~~~~~~~~~~

In the main text, we have identified black funnels and black droplets in the AdS C-metrics for $\lambda > -1$.  While taking $\lambda = -1$ with finite $\ell$ yields the flat-space C-metric, one might wonder if new asymptotically AdS metrics might be obtained by taking $\lambda \to -1$ holding $\ell_4$ fixed instead. Since this requires $\ell \to 0$, we must also scale $x$ and $y$ to obtain a finite limiting metric. This scaling means that we effectively zoom in on some point $(x_+,y_+)$ in $x,y$ space. For $x_+ \neq y_+$, one obtains only pieces of flat Minkowski space.  We therefore focus on the case $x_+ = y_+$ below.

Note that
for $\lambda =-1$ we have $F = -G$, so that the roots $F$ coincide in this limit with the ($\lambda$-independent) roots of $G$.  When $x_+$ is not a root of $G$, the non-trivial scaling limit leads to pure AdS space in a slightly twisted version of Poincar\'e coordinates. However, the behavior is more interesting when $x_+$ is a root of $G$.  Since triple roots of $G$ do not arise, there are only two cases to consider.  As we will see, the scaling limit is independent of all parameters and cares only about the degree of the root $x_+$.

 \paragraph{Case 1:} $x_+$ is a single root of $G$. \\
Consider the scaling limit
\begin{eqnarray}
&&\lambda \to -1 \ , \qquad \text{with} \nonumber \\
&&X = \frac{x-x_+}{\lambda +1} \ , \qquad Y = \frac{y-x_+}{\lambda +1} \ , \qquad \frac{\ell}{ \sqrt{1+\lambda }} \ , t, \;\; \phi \ ,\qquad \text{fixed}
\label{}
\end{eqnarray}	
Denoting $G'(x_+)=  G_+$, we
find that the metric \req{adsc} reduces to
\begin{equation}
ds^2 = \frac{\ell_4^2}{(X-Y)^2} \, \left(-G_+\, Y \, dt^2 + \frac{dY^2}{G_+ \, Y} + \frac{dX^2}{1- G_+\, X} +(1-G_+ \, X) \, d\phi^2 \right).
\label{case1}
\end{equation}	
We take $G_+ > 0$ without loss of generality, using the freedom to redefine $(X,Y) \to (-X,-Y)$ to change the sign of $G_+$ if necessary.  The result \req{case1} can then be simplified using the coordinate change $\rho^2 = 1-G_+\, X$, $\zeta^2 = G_+ \, Y$ to yield
\begin{eqnarray}
ds^2 &=& \frac{\ell_4^2\, G_+^2}{(\rho^2+\zeta^2-1)^2} \, \left( -\zeta ^2\, dt^2 + \frac{4\, d\zeta^2}{G_+^2} + \frac{4 \, d\rho^2}{G_+^2} + \rho^2\,d\phi^2\right) \nonumber \\
&=& \frac{4\, \ell_4^2}{(\xi^2-1)^2}\, \left( -\xi^2\,\sin^2\theta\,  dT^2 + \xi^2\, \cos^2 \theta\, d\Phi^2 + d\xi^2 + \xi^2 \, d\theta^2\right),
\label{}
\end{eqnarray}	
where we have made some additional trivial coordinate changes. This is the metric of \AdS{4} in the dS$_3$ slicing.  The boundary is at $\xi =1$ and we clearly see dS$_3$ in static coordinates. We can furthermore check that this solution has no stress tensor by passing to the Fefferman-Graham coordinate chart using the transformation
\begin{equation}
\xi = \frac{1-z^2}{1+z^2}
\label{}
\end{equation}	
to write the metric as
\begin{equation}
ds^2 = \frac{4\,\ell_4^2}{z^2}\, \left(dz^2 + \left(\frac{1-z^2}{1+z^2}\right)^2\, ds_{dS_{3}}^2\right).
\label{}
\end{equation}	
The absence of odd powers in the small $z$ (near boundary expansion) implies that $T_{\mu \nu}\equiv 0$.

\paragraph{Case 2:} $x_+$ is a double root of $G$. \\
This case requires $\kappa=1$ and $\mu = \mu_c$, and so was discussed in \cite{Hubeny:2009hr}. Briefly, the limit
\begin{eqnarray}
&&\lambda \to -1 , \;\;\mu = \mu_c  , \qquad \text{with} \nonumber \\
&&X = \frac{x-x_0}{\sqrt{\lambda +1}} \ , \quad Y = \frac{y-x_0}{\sqrt{\lambda +1}} \ , \quad \frac{\ell}{ \sqrt{1+\lambda }} \ , \quad \Phi = \sqrt{1+\lambda}\, \phi , \quad t, \quad \text{fixed} .
\label{}
\end{eqnarray}	
leads to the simple metric
\begin{equation}
ds^2  = \frac{\ell_4^2}{(X-Y)^2} \, \left( -(1-Y^2)\,dt^2  + \frac{dY^2}{1-Y^2} +\frac{dX^2}{X^2} + X^2 \, d\Phi^2\right).
\label{adsclm1m}
\end{equation}	
We have a horizon at $Y = \pm 1$ and infinities at $X = 0$, $X = \pm \infty$.  Note that all singularities have disappeared.  The induced metric on the boundary \req{indbdymet} becomes
\begin{equation}
ds^2 = -(1-X^2)\, dt^2 +\frac{dX^2}{X^2\,(1-X^2)} + X^2\, d\Phi^2.
\label{}
\end{equation}	

While there are no rotation axes, we can apply the definitions of black funnels and black droplets given in \sec{s:coordcmet} if replace the axes at $x_0,x_1,x_2$ by the infinities $X=0,X = \pm \infty.$ Due to the symmetry under $(X,Y) \to (-X,-Y)$, there are only two distinct choices of coordinate domains.  For $Y < X < 0$, the horizon at $Y=-1$ is a black funnel.    For $Y < X, X > 0$, the horizon at $Y=1$ is a (non-compact version of) a black droplet suspended above a planar black hole at $Y = -1$.

As for the case of a single root, the boundary stress tensor \req{Tupdn} vanishes.   To gain some perspective on this statement, recall from \cite{Hubeny:2009hr} that (\ref{adsclm1m}) asymptotes near $X=0$ to the $M=0$ black hole of \cite{Emparan:1999gf} which, although it describes a deconfined phase, also has $T_{\mu \nu} =0$.  The vanishing of $T_{\mu \nu}$ is due to a precise cancelation between the stress-energy of the deconfined plasma in this state and the stress tensor induced by vacuum polarization.

%%%%%%%%%%%%%%%%%%%%%%%%%%%%%%%%%%%%%%%%%%%%
%\bibliography{bfunnel}
%\bibliographystyle{utphys}

\providecommand{\href}[2]{#2}\begingroup\raggedright\endgroup

\end{document}